\documentclass[prb,a4paper,twocolumn,superscriptaddress,showpacs,amsfonts,amssymb,amsmath, floatfix]
{revtex4-1}
\usepackage{graphicx}
\usepackage{tabularx}
\usepackage{subfigure}
\mathchardef\mhyphen="2D
\usepackage{times}
\usepackage[usenames,dvipsnames]{color}

\begin{document}

\author{Sananda Biswas}  \affiliation{Theoretical Sciences Unit, Jawaharlal Nehru Centre for Advanced Scientific Research, Jakkur, Bangalore, India}
\author{Gustav Bihlmayer} \affiliation{Peter Gr\"unberg Institut and Institute for Advanced Simulation, Forschungszentrum J\"{u}lich and JARA, 52425 J\"ulich, Germany}
\author{Shobhana Narasimhan}  \affiliation{Theoretical Sciences Unit, Jawaharlal Nehru Centre for Advanced Scientific Research, Jakkur, Bangalore, India}\affiliation{Sheikh Saqr Laboratory, ICMS, Jawaharlal Nehru Centre for Advanced Scientific Research, Jakkur, Bangalore, India}\
\author{Stefan Bl\"ugel} \affiliation{Peter Gr\"unberg Institut and Institute for Advanced Simulation, Forschungszentrum J\"{u}lich and JARA, 52425 J\"ulich, Germany}
\date{\today}

\begin{abstract}
We have used {\it ab initio} density functional theory to compute the magnetic ground states of 
the surface alloy systems FeAu$_2$/Ru(0001) and MnAu$_2$/Ru(0001). For both systems, we find that 
the lowest energy magnetic configuration corresponds to a left-rotating spin spiral, in which the 
sense of rotation is determined by the Dzyaloshinskii-Moriya interaction.
These spirals are lower in energy than the ferromagnetic 
configuration by 3--4 meV per nm$^2$. We also find that FeAu$_2$/Ru(0001) has a significantly high 
magnetic anisotropy energy, of the order 1 meV per Fe atom. By comparing with the corresponding 
freestanding alloy monolayers, we find that the presence of the Ru substrate plays a significant 
role in determining the magnetic properties of the surface alloy systems.
   
\end{abstract}

\pacs{75.70.-i, 75.70.Ak, 75.25.-j}

\title{Spin Spirals in Surface Alloys on Ru(0001): A First-principles Study }

\maketitle

\section{Introduction}

\textit{}

The spin-orbit interaction in magnetic systems leads to the possibility of stabilizing exotic non-collinear magnetic
structures, such as spin spirals induced by the Dzyaloshinskii-Moriya (DM) interaction.\cite{dzyaloshinskiiJETP1957, moriyaPR1960} 
The recent emergence of the field of spintronics has led to additional interest in such systems. As an example, it has 
been suggested that the spin torque arising from the flow of a spin-polarized current through a system where the spins 
are arranged in a chiral fashion can lead to various phenomena such as the switching of magnetization, and microwave 
emission.\cite{kiselevNature2003,krivorotovScience2005} The Spin-orbit interaction can also lead to obtaining magnetic 
structures with significantly enhanced magnetic anisotropy energy (MAE).\cite{vanvleckPR1937} The MAE is defined as the 
energy barrier that has to be overcome to orient the magnetization oppositely, along the easy axis, by rotating it through 
a hard axis; for information storage applications, it is crucial that it should have a high value, so that stored data is 
not lost through thermal fluctuations. In this work, we use {\it ab initio} density functional theory to explore such issues 
for ultrathin surface alloys, created by the co-deposition of two bulk-immiscible elements on a substrate.

In a spin spiral, the moments on all magnetic atoms have approximately the same magnitude, but their direction rotates by a 
phase factor as one proceeds from one atom to the next, along the direction of propagation. Only a few thin-film systems have 
been shown to display a spin spiral ground state: single and double layers of Mn on W(110),\cite{bodeNature2007, yoshidaPRL2012} 
a single monolayer of Mn on W(001),\cite{ferrianiPRL2008} a double monolayer of Fe on W(110),\cite{mecklerPRL2009} and a PdFe 
bilayer on Ir(111).\cite{rommingScience2013} It is therefore appealing to see whether alloying can lead to more systems of this 
kind. 

Surface alloys are systems where two or more elements form an alloy restricted to the surface layer alone. Recent interest in 
such systems has been particularly boosted by the finding that it is possible to form surface alloys out of elements that are 
immiscible in the bulk.\cite{nielsenPRL1993, madhuraPRB2009} In early work, it was believed that the driving force for alloy 
formation in such systems was primarily the relief of surface stress.\cite{tersoffPRL1995} However, more recently it has been 
shown that when one of the constituents is a magnetic element, the dominant effect in driving mixing can, in some cases, be 
magnetism.\cite{bluegelAPA1996, mehendalePRL2010}

In this study, we focus on two supported systems, FeAu$_2$/Ru(0001) and MnAu$_2$/Ru(0001). In order to better distinguish those 
effects that arise from the presence of the Ru substrate, we also consider two hypothetical systems, viz., a freestanding FeAu$_2$ 
monolayer, and a freestanding MnAu$_2$ monolayer, both maintained at a nearest-neighbor spacing equal to that in bulk Ru. Ru 
crystallizes in the hexagonal close packed (hcp) structure, and thus the Ru(0001) surface has a triangular lattice, which offers 
an ideal substrate to study magnetic frustration, which can lead to a variety of interesting magnetic structures. Fe and Mn are 
both magnetic elements, but while bulk Fe is ferromagnetic, bulk $\alpha$-Mn is non-collinear antiferromagnetic at room 
temperature.\cite{yamadaJPSJ1970} Moreover, while Fe and Au are bulk-immiscible, Mn and Au form bulk alloys, such as MnAu$_2$, which 
displays a helical arrangement of the spins on Mn atoms.\cite{harpinJPR1961, udvardiPRB2006} When a  monolayer of Fe is deposited on 
Ru(0001), the resulting Fe/Ru(0001) system has a 120$^{\circ}$ N\'{e}el state.\cite{hardratPRB2009} In contrast, Mn/Ru(0001) displays 
a row-wise antiferromagnetic structure.\cite{dupuisPRB1993} FeAu$_2$/Ru(0001) has been shown, both experimentally and 
theoretically,\cite{mehendalePRL2010} to have a pseudomorphic $(\sqrt 3 \times \sqrt 3)$ structure, with long-range-order. In this 
structure, the Fe atoms in the overlayer constitute a triangular superlattice, and every Fe atom is surrounded by six Au atoms [see 
Fig.~\ref{fig:system_top_view}]. It has been shown, by density functional theory calculations, that this structure is stabilized primarily 
by magnetism rather than stress relief.\cite{mehendalePRL2010} MnAu$_2$/Ru(0001) has a similar structure, with the Fe atoms replaced 
by Mn atoms. We have found that this structure is also stable against phase-segregation.\cite{biswas_paper} The presence of Au and Ru 
atoms is interesting for our purpose, since they are expected to enhance spin-orbit coupling, and thus increase both the DM interaction 
and the MAE.

The outline of this paper is as follows. In Section II, we describe the four systems on which we have carried out our calculations. Next, 
in Section IIIA, we lay out the relevant formalism, describing separately each of the three main contributions to the total energy, viz., 
the symmetric Heisenberg exchange energy, the Dzyaloshinskii-Moriya energy, and the magnetic anisotropy energy. Next, in Section IIIB, we 
give the technical details of our calculations. Our results are presented in Section IV. Section V contains a discussion of our results. 
We present a summary in Section VI. In an appendix, we discuss issues related to the applicability of the force theorem.

\section{Systems}

As mentioned above, in order to clearly separate out the effects of the substrate, we perform calculations on \textit{X}Au$_2$ layers 
(\textit{X} = Fe or Mn), both with and without the Ru(0001) substrate. Thus, we have studied four systems: (A) freestanding FeAu$_2$ 
monolayer, (B) freestanding MnAu$_2$ monolayer, (C) FeAu$_2$/Ru(0001), and (D) MnAu$_2$/Ru(0001). Note that in all four cases, the 
in-plane nearest-neighbor spacing is fixed as equal to the experimental value for Ru(0001) $=$ 2.70 \AA.

\begin{figure} [h]
\centering
\subfigure[]{\label{fig:system_top_view} \includegraphics[width=3cm]{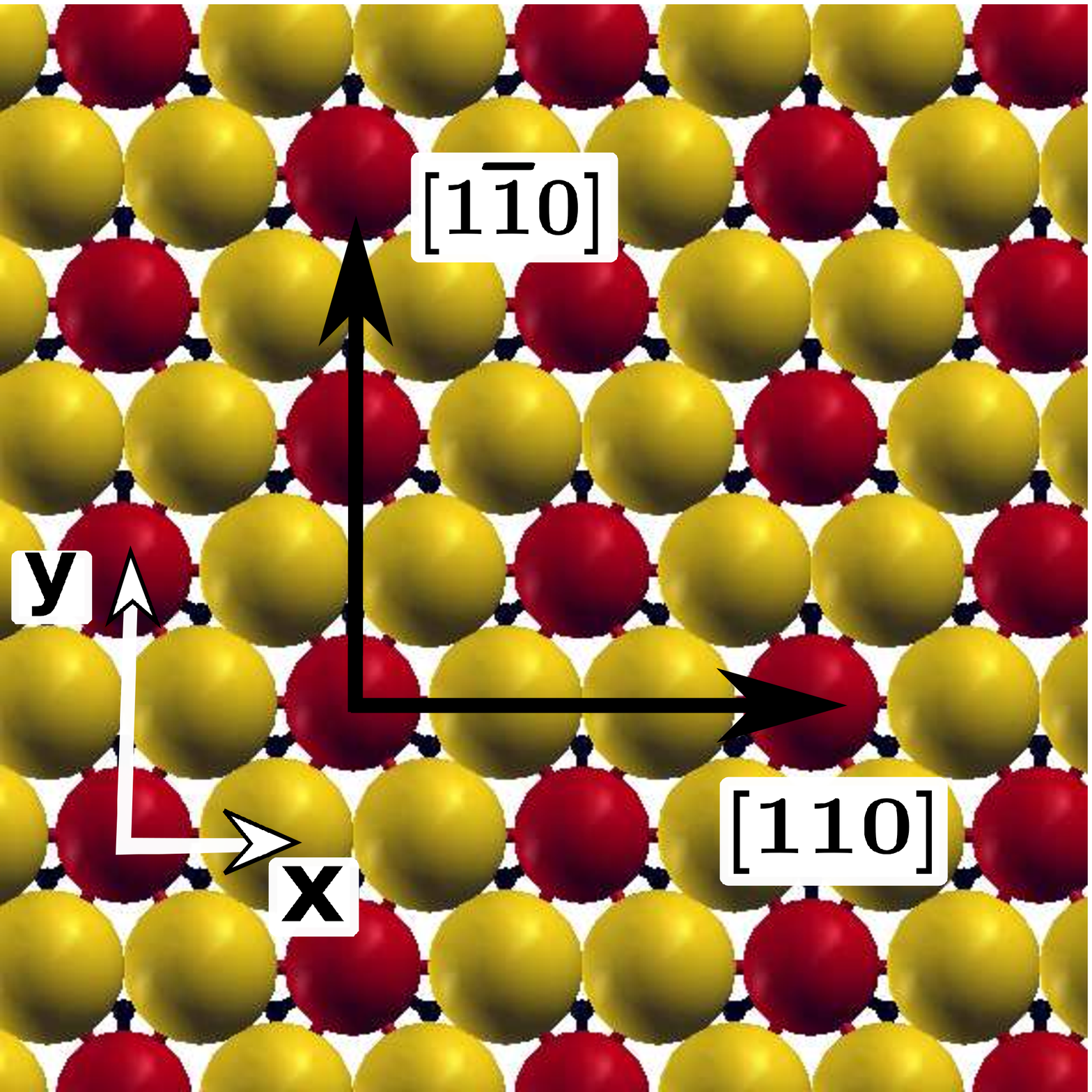}}
\hspace{10pt}
\subfigure[]{\label{fig:k_space}  \includegraphics[width=2.6cm]{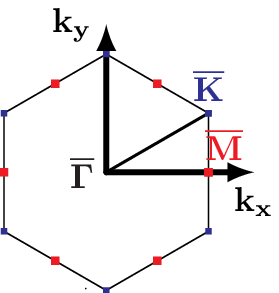}
}
\caption{(Color online) System geometry in real and reciprocal space: (a) shows the top  view of the alloy 
monolayer, \textit{X}Au$_2$, (\textit{X} $=$ Fe or Mn) on Ru(0001). The dark (red) and light (yellow) spheres 
represent  \textit{X} and Au atoms, respectively. The black dots indicate the topmost layer 
of the Ru atoms, for those systems in which the Ru substrate is present. 
(b) shows the corresponding hexagonal 
surface Brillouin zone and the high symmetry points $\overline{\Gamma}$, $\overline{\mathrm{M}}$, and 
$\overline{\mathrm{K}}$.}
\label{fig:system_image}
\end{figure}

Fig.~\ref{fig:system_top_view} shows the top view of all four systems.  The corresponding  surface Brillouin zone, 
along with high symmetry points, is shown in 
Fig.~\ref{fig:k_space}. When computing spin spirals with 
wavevector $\mathbf{q}$, the $\overline{\Gamma}$ point (zone center) 
corresponds to the ferromagnetic state,  $\overline{\rm{K}}$ corresponds to a row-wise antiferromagnetic 
state, and $\overline{\rm{M}}$ corresponds to a 120$^\circ$ N\'{e}el state. Points in the interior of the 
Brillouin zone correspond to general spin spirals.

\section{Method}

\subsection{Formalism}

For a spin spiral with wavevector $\mathbf{q}$, the magnetic moment of
a magnetic atom at position $\mathbf{R}$ is given by: 

\begin{equation}\label{eqn:mag_mom_q}
\mathbf m(\mathbf R) = m
\begin{pmatrix}
  {\rm sin} \alpha \ {\rm cos} (\mathbf{q} \cdot \mathbf{R})\\
  {\rm sin} \alpha \ {\rm sin} (\mathbf{q} \cdot \mathbf{R})\\
  {\rm cos} \alpha 
 \end{pmatrix},
\end{equation}

\noindent
where $m$ is the magnitude of the magnetic moment, and $\alpha$ is the cone angle of the spin spiral. 
The total energy of such a spin spiral is given by the sum of three terms:

\begin{equation} \label{eqn:energy_total}
E_{\rm{total}} (\mathbf{q}) = E_{\rm{HE}}({\mathbf{q}})  + E_{\rm{DM}}({\mathbf{q}}) + K_{\rm{avg}}.
\end{equation}

\noindent
In Eq.~(\ref{eqn:energy_total}), the first term on the right-hand-side, $E_{\rm{HE}}$, is the energy due to 
symmetric Heisenberg exchange. The second term, $E_{\rm{DM}}$, arises from the Dzyaloshinskii-Moriya 
interaction.  The third term, $K_{\rm{avg}}$, denotes the average value of the MAE over one wavelength 
$\lambda$ of the spin spiral. Note that in the absence of spin-orbit coupling, only the first of these 
three terms would be present (magnetic dipole-dipole interactions in low-dimensional systems generally being
very weak). Throughout this paper, the zero of energy is defined such that $ E_{\rm{HE}}(\mathbf{q}=0)=0$. 
As $|\mathbf{q}|=2 \pi/\lambda$, we can also write Eq.~(\ref{eqn:energy_total}) as a function of $\lambda^{-1}$.

In this paper, we restrict ourselves to considering only homogeneous, cycloidal, planar spin spirals. In 
homogeneous spin spirals the relative angle between neighboring 
spins is always a constant. In order to describe a spin spiral, in addition to a wavevector $\mathbf{q}$, one 
must specify an axis about which the spins rotate. In helical spin spirals, this axis of rotation is parallel 
to $\mathbf{q}$, whereas in cycloidal spin spirals, it is perpendicular to $\mathbf{q}$. Symmetry arguments predict 
that, on an isotropic surface, cycloidal spin spirals will always be lower in energy than helical 
spin spirals.\cite{ferrianiPRL2008, crepieuxJMMM1998} In planar spin spirals, the spins are confined to a plane 
normal to the rotation-axis;  the  Dzyaloshinskii-Moriya interaction is expected to be largest in such a 
situation.\cite{dzyaloshinskiiJETP1957, izyumovSPU1984}         

Below, we describe each of the terms contributing to the total energy in Eq.~(\ref{eqn:energy_total}).

\subsubsection{Heisenberg Exchange Energy $E_{\rm{HE}}$}

If we consider a system consisting of spins $\{\mathbf{S}_{i}\}$, on lattice sites $i$, then one can write 
the contribution to the total energy from the Heisenberg exchange interaction as:

\begin{equation} \label{eq:heisenberg_eq}
E_{\rm{HE}} =  -\displaystyle \sum_{i < j}  J_{ij} ({\bf S}_i \cdot {\bf S}_j),
\end{equation}

\noindent
where $J_{ij}$ is the exchange integral and the sum runs over all pairs of distinct lattice sites $i$ and 
$j$. Note that this interaction is symmetric with respect to exchange of spins between sites. 

To calculate $E_{\rm{HE}}$ for a spin spiral with wavevector $\mathbf{q} \ne 0$ requires, in principle, the
use of a supercell. This would hugely increase the computational time, especially for spin spirals of long 
wavelength. However, the use of supercells can be avoided by making use of the generalized Bloch 
theorem;\cite{sandratskiipssb1986} this permits one to carry out all calculations making use of the 
chemical unit cell. 

There are two possible approaches for calculating $E_{\rm{HE}}$. The quicker, but less accurate way, 
is to make use of Andersen's force theorem, also referred to as the magnetic force theorem or frozen 
force theorem.\cite{mackintosh1980} This states that the change in energy due to the presence of a small 
perturbation can be calculated non-self-consistently from the eigenvalue sum, if the self-consistent 
solution of the unperturbed Hamiltonian is known. It is generally assumed that the force theorem can be 
applied for most small perturbations, and can be used, e.g., to  calculate the energy difference 
$\delta E_{\rm{HE}}$ between two spin spirals of slightly different wavelengths, or the MAE. 

The more time-consuming, but also more accurate, approach for calculating $E_{\rm{HE}}$ is to perform 
fully self-consistent calculations. In this approach, the ground state electronic and magnetic densities 
($n_{0}$, $\mathbf{m}_{0}$) are calculated self-consistently for each spin spiral, so as to yield a 
precise value for the energy difference $\delta E_{\rm{HE}}$ between two spin spirals of different wavelengths.

Given the wide use of the force theorem in calculations of magnetic structure, it would be of interest to 
obtain some insight into its domain of applicability, and to examine how accurate results obtained using it, 
are. For this reason, we have used both the force theorem and self-consistent approaches in this paper, and 
present a comparison of results obtained using the two techniques.

\subsubsection{Dzyaloshinskii-Moriya Energy $E_{\rm DM}$}

In an inversion-asymmetric system, such as a surface, not only the symmetric Heisenberg exchange 
interaction, but also the antisymmetric exchange (DM) interaction becomes important, and can play a crucial 
role in determining the magnetic ground state of the system.\cite{dzyaloshinskiiJETP1957, moriyaPR1960} 
The energy due to the DM interaction can be written as:

\begin{equation} \label{eqn:DMI}
E_{\rm{DM}} =  \displaystyle \sum_{i < j} {\bf D}_{ij} \cdot ({\bf S}_i \times {\bf S}_j),
\end{equation}

\noindent
where ${\mathbf D}_{ij}$ is the DM vector, and the sum again runs over distinct lattice sites. It has been 
shown that depending on the symmetry of a system, some or all components of $\mathbf{D}$ may 
vanish.\cite{crepieuxJMMM1998} The non-zero components of ${\mathbf{D}}$ can be obtained by spin spiral
calculations varying $\mathbf{q}$ along different crystallographic directions. For planar cycloidal spin 
spirals on a surface, such as those considered in this study, the component of ${\mathbf{D}}$ along 
$\mathbf{q}$ vanishes by symmetry. However, non-zero components of ${\mathbf{D}}$, which are orthogonal to 
$\mathbf{q}$, may exist. For example, if $\mathbf{q}$ lies along the $x$-axis ([110] direction), the non-zero 
components of ${\mathbf{D}}$ can be $D_y$ and $D_z$, whereas, if $\mathbf{q}$ lies along the $y$-axis 
([1$\overline{1}$0] direction), then the non-zero components can be $D_x$ and $D_z$ [see Fig.~\ref{fig:system_top_view}].  
Note that we find that the freestanding FeAu$_2$ and MnAu$_2$ monolayers remain completely flat,  i.e., display no buckling, 
and thus, by symmetry, the DM interaction is absent for these systems.

To obtain $E_{\rm{DM}}$, one can solve the Dirac equation self-consistently. To do this one would, in principle, 
need to use large supercells, as the generalized Bloch theorem breaks down in the presence of spin-orbit coupling. 
However, to deal with this problem, a technique has been developed \cite{heidePBCM2009} to obtain $E_{\rm DM}$ 
within the chemical unit cell, treating the spin-orbit coupling as a small perturbation to first order. We have 
employed this method for calculating $E_{\rm{DM}}$.

\subsubsection{Magnetic Anisotropy Energy $K$}

The magnetic anisotropy energy (MAE) is the height of the energy barrier that has to be overcome to reverse the 
direction of the spin along the easy axis, and is  given by:

\begin{equation}  \label{eqn:mae_soc}
K = E{^{\rm {hard \mhyphen axis}}} - E{^{\rm {easy \mhyphen axis}}},
\end{equation} \date{\today}

\noindent
where $E{^{\rm {hard \mhyphen axis}}}$  and $E{^{\rm {easy \mhyphen axis}}}$ are the total energies of the 
system with magnetization along the hard$\mhyphen$axis and the easy$\mhyphen$axis, respectively, in the plane 
of rotation of the spins. The MAE ($K$) has two contributions, $K_{\rm{SO}}$ and $K_{\rm{dip}}$, 
arising from the spin-orbit (SO) coupling and the magnetic dipole-dipole interaction, respectively.

One can perform either self-consistent calculations or use the force theorem to obtain the value of $K$. We 
have used both methods and compared the results.

\subsection{Calculation Details}

We have used density functional theory (DFT) as implemented in the {\sc Fleur} code,\cite{fleur} which is based on 
the Full-potential Linearized Augmented Plane-wave (FLAPW) method. Exchange-correlation interactions were 
treated using a generalized gradient approximation of the Perdew-Burke-Ernzerhof form.\cite{PBE} The muffin-tin 
radii of Mn, Fe, Au and Ru were set equal to 2.56, 2.32, 2.42, and 2.32 a.u., respectively. The cutoff for the 
$\ell$-value of the basis set consisting of spherical harmonics was fixed at 12, in order to expand the 
wavefunction inside the muffin-tins,  while the $\ell$-cutoff for the non-spherical part of the Hamiltonian 
was chosen to be 8. The plane-wave cutoff for the basis set used to expand the electronic wavefunction in the 
interstitial region was 3.6 a.u.$^{-1}$; this was increased to 4 a.u.$^{-1}$ when calculating the MAE, in order to achieve the 
increased accuracy necessary here. For the charge density and the exchange-correlation part of the potential, 
the plane-wave cutoffs were 12.3 a.u.$^{-1}$ and 10.3  a.u.$^{-1}$, respectively. 
      
We have 
considered two collinear magnetic configurations: ferromagnetic (FM) and row-wise AFM. For the 
freestanding \textit{X}Au$_2$ monolayer systems, the chemical (primitive) unit cell contains three 
atoms: one \textit{X} atom and two Au atoms. All the calculations for the \textit{X}Au$_2$ alloy monolayers 
were carried out within this unit cell, except while performing collinear magnetic calculations for the
antiferromagnetic (AFM) state. For these calculations a rectangular supercell, containing two \textit{X} atoms and 
four Au atoms, was used.  All the atoms were relaxed until the forces on each atom were less than 1 mHa/a.u. 
In both the FM and row-wise AFM configurations, we found that the freestanding  FeAu$_2$ and MnAu$_2$ monolayers remained 
completely flat upon relaxation, i.e., no buckling was observed.

For the \textit{X}Au$_2$/Ru(0001) systems, the Ru(0001) substrate was modeled by a slab containing six atomic 
layers of Ru. For geometric optimization and MAE calculations, a symmetric slab was used, in which the alloy 
monolayer was placed on both sides of the slab. When optimizing geometries, the alloy layers and the first two Ru
layers were allowed to relax, with the 
same force convergence criterion as used for the freestanding monolayers. To calculate $E_{\rm{HE}}$ and 
$E_{\rm{DM}}$ in an inversion-asymmetric environment, an asymmetric slab was used, in which the alloy monolayer 
was deposited  on only the upper surface of the six-layer Ru slab. 

We found that for \textit{X}Au$_2$/Ru(0001), the atoms on the overlayer buckled upon relaxation, in both FM and 
row-wise AFM configurations. Being larger, the Au atoms relax further away from the substrate,
while the smaller $X$ atoms remain closer to the substrate; the degree of buckling is quite significant. In order to 
quantify the degree of buckling, we computed $d_{\rm {b}}= d{_{\rm {Au}}}-d{_{X}}$, where $d{_{\rm{Au}}}$ and $ d{_{X}}$ 
are the $z$ coordinates of the Au and $\textit{X}$ atoms, respectively. We found that for 
FeAu$_2$/Ru(0001), $d_{\rm {b}}= 0.38$ \AA\ for both FM and AFM configurations, while for MnAu$_2$/Ru(0001),
the values of $d_{\rm {b}}$ were 0.22 \AA\ and  0.27 \AA\ in the FM and row-wise AFM configurations, respectively. This buckling
plays an important role in the DM interaction, as we will see further below.

The interlayer distance along the $z$-direction between the $X$ atom and the first Ru layer is 2.12 \AA\ for FeAu$_2$/Ru(0001)
and 2.30 \AA\ for MnAu$_2$/Ru(0001). The larger value in the latter case can be attributed to the presence of larger 
magnetic moments on Mn atoms than on Fe atoms. 

The different contributions to the total energy in Eq.~(\ref{eqn:energy_total}) differ in magnitude, and thus 
require differing degrees of accuracy. For this reason, we have separately checked the convergence of each of 
these contributions, with respect to the density of Brillouin zone (k-point) sampling.  In all cases, a smearing
of width 0.001 Ha was used to improve the convergence, except for the calculation of MAE, where a smaller smearing 
width of 0.0001 Ha was used.

\section{Results}

We first perform collinear magnetic calculations; these are useful not only because they might correspond
to the magnetic ground state, but because the relative energies of FM and AFM states help to gauge
the likelihood of obtaining non-collinear states such as spin spirals. We then go on to perform calculations
on spin spirals with wavevectors $\mathbf{q}$ along high-symmetry directions of the Brillouin zone.

\subsection{Collinear Magnetic Structures}

We have considered two collinear magnetic configurations: (i) 
FM and (ii) row-wise AFM, for all four systems under study. These calculations were 
carried out using 132 $\mathbf{k}_{\|}$-points in the irreducible part of the surface Brillouin zone for FeAu$_2$/Ru(0001) 
and MnAu$_2$/Ru(0001), while for FeAu$_2$ and MnAu$_2$ monolayers 128 $\mathbf{k}_{\|}$-points were used in the irreducible 
Brillouin zone. 

For the freestanding FeAu$_2$ monolayer, we find that the FM configuration is lower in energy than the AFM 
configuration, whereas for the freestanding MnAu$_2$ monolayer, the reverse is true. The energy difference 
between the two collinear magnetic structures considered, $\Delta E_{\mathrm{AFM-FM}}$, is  64 meV per Fe 
atom, for FeAu$_2$, and  $-70$ meV per Mn atom for MnAu$_2$. However, when deposited on the Ru(0001) substrate, 
the ferromagnetic state is lower in energy for both the Fe and Mn surface alloys; the value of 
$\Delta E_{\mathrm{AFM-FM}}$ is found to be 62 meV per Fe atom for FeAu$_2$/Ru(0001), and  19 meV per Mn atom 
for  MnAu$_2$/Ru(0001). The fact that the stability of the magnetic structure switches from being row-wise AFM for the 
freestanding monolayer, to FM for the deposited monolayer, for MnAu$_2$, is already an indication that the 
presence of the Ru substrate can play an important role in determining the magnetic properties of the system. 

\begin{table}[!ht]
\centering
\caption{The magnetic moments on the various atoms in the alloy layer, and the top
two layers of the substrate (where present), for the collinear ground state of the four systems studied. Ru(1) and Ru(2) atoms 
lie in the first and second layer, respectively, of the Ru substrate. See the text for the description 
of Ru(2)-I and Ru(2)-II. 
}
\begin{tabular}{l r r r r}
\hline 
\hline
Atom         &  \multicolumn{4}{c}{Magnetic moments ($\mu_B$)}\\
\cline{2-5}
             & FeAu$_2$/Ru  & FeAu$_2$  & MnAu$_2$/Ru & MnAu$_2$ \\  
\hline
Fe/Mn     &  2.88   & 3.2  &  3.75 & $\pm 4.14$  \\
Au        &  0.02   & 0.04 &  0.02 & $\pm 0.02$  \\
Ru(1)     &  0.00   & -    & $-0.09$ &  -         \\
Ru(2)-I   & $-0.03$ & -    & $-0.02$ &  -         \\
Ru(2)-II  & $-0.11$ & -    & $-0.07$ &            \\ 
\hline
\hline
\end{tabular}
\label{table:mag_moms}
\end{table}

In Table~\ref{table:mag_moms}, we have listed the magnetic moments for the different types of atoms in the 
row-wise AFM (for MnAu$_2$ monolayers) and FM (for the other three systems) configurations. As expected, the 
magnetic moments of the Fe and Mn atoms are higher in the freestanding monolayers, where the atoms have a 
lower coordination than when they are deposited on the Ru substrate. 
We find that the magnetic moments on the Au atoms tend to be aligned parallel to the $X$ atoms, implying a ferromagnetic 
interaction between them, whereas, in general, the magnetic moments on the Ru atoms in the substrate are aligned
opposite to those of the $X$ atoms. Note that there are two inequivalent types of Ru atoms in the second layer 
of the substrate, labeled as Ru(2)-I and Ru(2)-II; these are situated directly below \textit{X} and Au atoms, 
respectively.

\subsection{Stability of the Surface Alloys}
\noindent

In order to check the stability of the surface alloys with respect to the phase segregated states 
[\textit{X}/Ru(0001) and Au/Ru(0001)], we have obtained $\Delta H$, the enthalpy of formation of the alloy, 
which is given by:

\begin{equation} \label{eqn:deltaH}
\Delta H = E_{X\rm{Au{_2}/Ru}} - \frac{1}{3} E_{X/\rm{Ru}} - \frac{2}{3} E_{\rm{Au/Ru}},
\end{equation}

\noindent
where $E_C$ is the total energy of the system $C$. 
The above equation applies in the presence of the Ru substrate; for the freestanding monolayers, we of course
use a similar equation involving freestanding systems with no Ru substrate present. Note that a positive/negative value of $\Delta H$ implies that the system is 
unstable/stable with respect to phase segregation. 

To enable us to gauge the contribution of exchange interactions to $\Delta H$, one can see how the enthalpy of
formation changes when spin polarization is suppressed. Thus, the first two terms on the
right-hand-side of Eq.~(\ref{eqn:deltaH}) are computed  in both the magnetic (M) and non-magnetic 
(NM) ground states; the corresponding values of $\Delta H$ are denoted as $\Delta H_{\rm{M}}$ and
$\Delta H_{\rm{NM}}$, respectively. Note that for these calculations the muffin-tin radius for the Mn atom 
has been taken to be 2.42 a.u. and 2.32 a.u. for the magnetic and non-magnetic calculations, respectively. 

The magnetic ground states used in computing $\Delta H_{\rm{M}}$ are as follows: for FeAu$_2$/Ru(0001)
and MnAu$_2$/Ru(0001), they are the ferromagnetic state obtained in section IV.A. For Fe/Ru(0001), it is a 
120$^\circ$ N\'{e}el state, which is lower in energy than the FM state by 58 meV per Fe atom.\cite{hardratPRB2009}
For Mn/Ru(0001), it is the row-wise AFM state. 
For the freestanding FeAu$_2$ and MnAu$_2$ monolayers, it is the FM and row-wise AFM states, 
respectively, obtained in this study. For the freestanding Fe monolayer, it is a 2Q state, \cite{al-zubi_pssb2011} 
and for a freestanding Mn monolayer it is the row-wise AFM state. 

\begin{table}[h!]
\centering
\caption{Listed below are the values of enthalpy of formation $\Delta H_{\rm{M}}$  and $\Delta H_{\rm{NM}}$ for the
magnetic and non-magnetic systems, respectively. Note that here all the systems are taken to be in their respective 
collinear magnetic ground states.}
\begin{tabular}{l l r r }
\hline 
\hline
System             & Collinear                 & $\Delta H_{\rm{NM}}$       & $\Delta H_{\rm{M}}$ \\
                   & magnetic state             &\multicolumn{2}{c}{(meV/surface atom)}\\
\hline
FeAu$_2$/Ru(0001)  & FM                         &  12    & $-154$ \\
MnAu$_2$/Ru(0001)  & FM                         &   7    & $-177$ \\       
FeAu$_2$           & FM                         &  31    & $-184$ \\
MnAu$_2$           & Row-wise AFM               &  73    & $-333$ \\
\hline
\hline
\end{tabular}
\label{table:delta_H}
\end{table}

The values of $\Delta H_{\rm{M}}$  and $\Delta H_{\rm{NM}}$ are listed in Table~\ref{table:delta_H}. 
Let us first focus on the values of $\Delta H_{\rm{NM}}$. These values are all positive, which implies that both the
freestanding and the supported alloys are unstable with respect to the phase segregated states of \textit{X}/Ru(0001) 
and Au/Ru(0001), when no magnetic interactions are present in the systems. However, if we focus on the values of 
$\Delta H_{\rm{M}}$, we find that the values become negative. This indicates that in the presence of magnetic interactions, 
all four surface alloys considered here become stable against phase segregation. Therefore, we conclude that magnetism plays 
a crucial role in the stability of these alloys.\cite{bluegelAPA1996, mehendalePRL2010, marathePRB2013}

In the phase-segregated situation, the magnetic moment in Fe/Ru(0001) is 2.75 $\mu_B$ per Fe atom;\cite{hardratPRB2009} 
this is increased to 2.88 $\mu_B$ in the surface alloy, where every Fe atom is surrounded by six Au atoms. This increase in 
magnetic moment and the magnetovolume effect provide the basic driving forces for the formation of the surface alloy. Similarly, 
the magnetic moment on the Mn atoms increases from 3.46 $\mu_B$ in Mn/Ru(0001) to 3.75 $\mu_B$ in MnAu$_2$/Ru(0001), which is why 
exchange interactions strongly favor the formation of the latter.

\subsection{Spin Spiral Calculations for \textit{X}Au$_2$/Ru(0001)}

We now proceed to the question of primary interest for us, viz., whether the two supported 
systems, FeAu$_2$/Ru(0001) and MnAu$_2$/Ru(0001), exhibit a spin spiral ground state. While 
doing this, we have made use of the optimized geometry obtained for the collinear ferromagnetic state. 
In presenting these results, we have separated out each contribution to the energy [see Eq.~(\ref{eqn:energy_total})] of the
spin spiral.

\subsubsection{Results for Heisenberg Exchange Energy $E_{\rm{HE}}$}

\begin{figure} [!h]
\centering
\includegraphics[width=8.0cm]{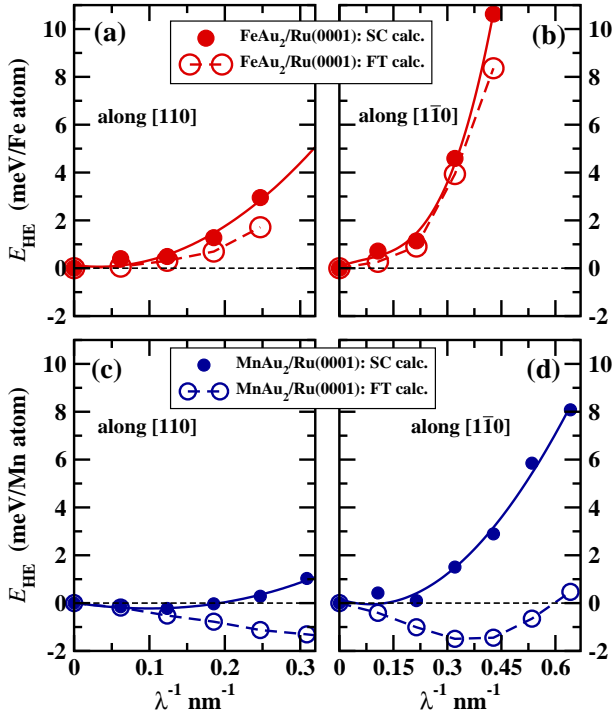}
\caption{(Color online) Dispersion of Heisenberg exchange energy $E_{\rm{HE}}$ for small $\lambda^{-1}$, for [(a) and (b)] 
FeAu$_2$/Ru(0001) and [(c) and (d)] MnAu$_2$/Ru(0001), along high-symmetry directions in the Brillouin zone.  
The open circles are the results from the force theorem (FT) calculations. The filled circles are results from the
self-consistent (SC) calculations and correspond to the zoomed-in regions of the top panel of Fig.~\ref{fig:ss_mag_mom}.}
\label{fig:ss_scf_ft}
\end{figure}

\begin{figure*} [!ht]
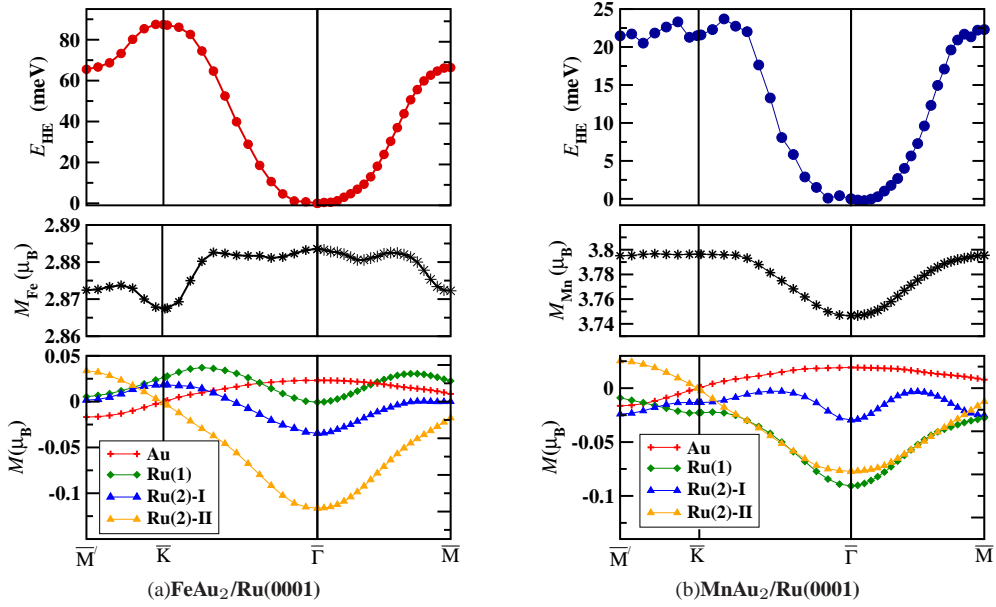

\centering
\subfigure[\bf{FeAu$_2$/Ru(0001)}]{\label{fig:ss_feau2_onRu_mag_mom} \includegraphics[width=6.0cm ]{Fig3a.eps}
}
\hspace{20pt}
\subfigure[\bf{MnAu$_2$/Ru(0001)}]{\label{fig:ss_mnau2_onRu_mag_mom} \includegraphics[width=6.0cm]{Fig3b.eps}
}\\
\hspace{8pt}
\caption{(Color online) Results from SC calculations for variation of Heisenberg exchange energy $E_{\rm {HE}}$, and 
magnetic moments $M$ on the different atoms, along high symmetry directions of the surface Brillouin zone, for (a) FeAu$_2$/Ru(0001) and (b) 
MnAu$_2$/Ru(0001). $E_{\rm {HE}}$ is given in meV per \textit{X} atom, and $M$ is given in $\mu_B$ per atom. See 
the text for the convention used in labeling atoms.}
\label{fig:ss_mag_mom}
\end{figure*}

As mentioned earlier, we have obtained $E_{\rm{HE}}(\lambda^{-1})$ using two possible approaches, the
force theorem (FT), and self-consistently (SC). Let us first consider the results obtained using the
former approach. In order to obtain converged results, we found that we need to use a very dense 
$\mathbf{k}_{\|}$-point mesh containing 6400 points in the full Brillouin zone. Since we know that the
FT approach should be valid only for small perturbations, we perturb about a SC solution corresponding 
to the FM state (i.e., $\bf {q} = 0$), and restrict ourselves to regions of the Brillouin zone in the 
vicinity of the zone center. In particular, we consider $|\lambda^{-1}| \le 0.32$  nm$^{-1}$  along the 
[110] direction, and $|\lambda^{-1}| \le 0.75$ nm$^{-1}$ along [1$\overline{1}$0]. Our results are shown 
by the open circles in Fig.~\ref{fig:ss_scf_ft}. [Note that as the relation 
$E_{\rm{HE}} (-\lambda^{-1}) = E_{\rm{HE}} (\lambda^{-1})$ holds for both systems, we have only shown the 
results for $\lambda^{-1} > 0$]. Interestingly, we find that for MnAu$_2$/Ru(0001), the graphs suggest 
that even with Heisenberg exchange interactions alone, a spin spiral state would be favored over the 
ferromagnetic state. However, for FeAu$_2$/Ru(0001), the ground state in this approximation remains the 
FM state.

Next, we proceed to verify these FT results by performing more accurate SC calculations. For these, we found
that it sufficed to use 512 and 800 $\mathbf{k}_{\|}$-points when sampling the irreducible Brillouin zones 
for FeAu$_2$/Ru(0001) and MnAu$_2$/Ru(0001), respectively. These results are shown by the filled circles in 
Fig.~\ref{fig:ss_scf_ft}. Somewhat surprisingly, the results obtained now are quite different, especially for 
MnAu$_2$/Ru(0001). The difference in energy between a spin spiral state and the FM state is now considerably 
reduced.

This suggests that results using the FT approach cannot always be trusted, and the force theorem must be
used with considerable caution. This point is discussed further in the Appendix to this paper.

We now go on to use the SC approach to compute $E_{\rm HE}$ throughout the Brillouin zone, along high-symmetry 
directions. This, along with results for the variation of magnetic moments, are shown in Fig.~\ref{fig:ss_mag_mom}.  
In this figure, $\overline{\Gamma}\overline{\rm{K}}$ and $\overline{\Gamma}\overline{\rm{M}}$ lie within the first 
Brillouin zone, while $\overline{\rm{K}}\overline{\rm{M'}}$ belongs to the second Brillouin zone. On examining 
this figure, we see that  for FeAu$_2$/Ru(0001), the lowest value of $E_{\rm{HE}}$ is at the $\rm{\overline {\Gamma}}$
point (see the top two panels of Fig.~\ref{fig:ss_mag_mom}). In contrast, for MnAu$_2$/Ru(0001), the lowest value of 
$E_{\rm{HE}}$ corresponds to a spin spiral with $\lambda^{-1} = 0.12 $ nm$^{-1}$ along the 
$\rm{\overline {\Gamma}\overline{M}}$ direction. This is more evident from Fig.~\ref{fig:ss_scf_ft}(c) (see the solid 
line and filled circles). Note however that: (i) the difference in $E_{\rm HE}$  between the FM state and the spin 
spiral state  is small, and (ii) to obtain the final result for ground state magnetic structure, we have yet to add 
the other two contributions $E_{\rm {DM}}$ and $K_{\rm {avg}}$, to $E_{\rm{HE}}$.  

The magnetic moments of the \textit{X} atoms, $M_X$, (shown by  the stars in the middle panel of 
Fig.~\ref{fig:ss_mag_mom}) vary only slightly as $\mathbf{q}$ changes. The induced moments on the Au, Ru(2)-I 
and Ru(2)-II atoms can be either positive (ferromagnetically aligned) or negative (antiferromagnetically aligned), 
depending on the value of $\mathbf{q}$; however the magnitude of these induced moments is small.

\subsubsection{Results for Dzyaloshinskii-Moriya Energy $E_{\rm{DM}}$}

\begin{figure} [!h]
\centering
\includegraphics[width=8.0cm ]{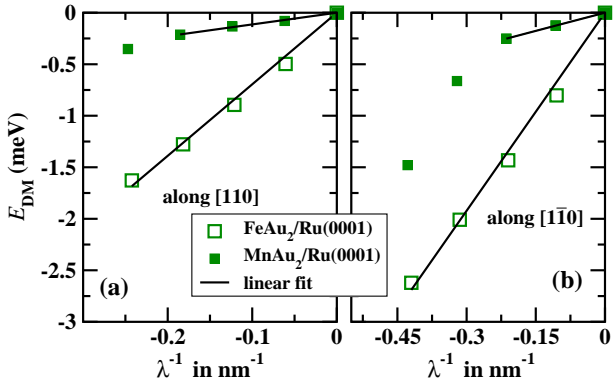}
\caption{(Color online) $ E_{\rm{DM}}$, contribution from the Dzyaloshinskii-Moriya interaction, to the total 
energy along (a) $[{110}]$  and (b) $[{1\overline{1}0}]$ for FeAu$_2$/Ru(0001) and MnAu$_2$/Ru(0001).}
\label{fig:dmi_onRu}
\end{figure}

We next calculate $E_{\rm{DM}}(\lambda^{-1})$ for FeAu$_2$/Ru(0001) and MnAu$_2$/Ru(0001). For all the 
spin spirals considered by us, the relation $E_{\rm{DM}}(- {\lambda^{-1}}) = - E_{\rm{DM}}({\lambda^{-1}})$ 
holds, where positive and negative values of $\lambda^{-1}$ correspond to right-rotating and left-rotating  
spirals, respectively. We have obtained the different components of ${\bf D}$ by varying $\bf{q}$ along 
the [110] direction, with $-0.25$ ${\rm nm}^{-1} < \lambda^{-1} < 0.25$ ${\rm nm}^{-1}$, and along the 
[1$\overline{1}$0] direction with $-0.42$ ${\rm nm}^{-1} < \lambda^{-1} < 0.42$ ${\rm nm}^{-1}$. The 
calculations were performed using 6400 $\mathbf{k}_{\|}$-points in the full surface Brillouin zone. Our results are presented 
in Fig.~\ref{fig:dmi_onRu}. For both FeAu$_2$/Ru(0001) and MnAu$_2$/Ru(0001), we have found that $E_{\rm{DM}}$ 
 always favors left-rotating spirals, we have therefore shown only the negative 
$\lambda^{-1}$ region in this figure. 

Our results for $E_{\rm{DM}}(\lambda^{-1})$ for FeAu$_2$/Ru(0001) and MnAu$_2$/Ru(0001) are shown by the open 
and filled squares, respectively, in Fig.~\ref{fig:dmi_onRu}. We find that the magnitude of $E_{\rm{DM}}$ is 
significantly larger for FeAu$_2$/Ru(0001) than it is for MnAu$_2$/Ru(0001) (the reason for this will be discussed
below); note that this was however also true of the magnitude of $E_{\rm{HE}}$, and that for negative 
$\lambda^{-1}$ these two terms have opposite sign, leading in both cases to a similar compensation of energies. 
Further, $E_{\rm{DM}}$($\lambda^{-1}$) is found to be linear for FeAu$_2$/Ru(0001), along  both  the [110] 
and  [1$\overline{1}$0] directions, for the  range of $\lambda^{-1}$ considered here. However, we can see that 
this is clearly not true for MnAu$_2$/Ru(0001), where $E_{\rm{DM}}(\lambda^{-1})$  is found to deviate from linear
behavior for $\lambda^{-1} \gtrsim -0.19$ nm$^{-1}$, along both directions.  In the region where $E_{\rm{DM}}$ varies 
linearly with $\lambda^{-1}$, we fitted our data to straight lines (see the black lines in Fig.~\ref{fig:dmi_onRu}), 
so as to obtain the components of $\bf{D}$ along different directions. We obtain $D_x$ and $D_y$ by fitting along 
the $[{1\overline{1}0}]$ and [110] directions, respectively; we find that the $D_z$ component always vanishes. 
For FeAu$_2$/Ru(0001), we obtain the values of $D_x$ and $D_y$ to be 6.40 and 6.94 meV nm$^2$, respectively, whereas 
for MnAu$_2$/Ru(0001), the values of $D_x$ and $D_y$ are found to be 1.17 and 1.12 meV nm$^2$, respectively. 

\begin{figure} [!h]
\centering
\includegraphics[width=7.5cm]{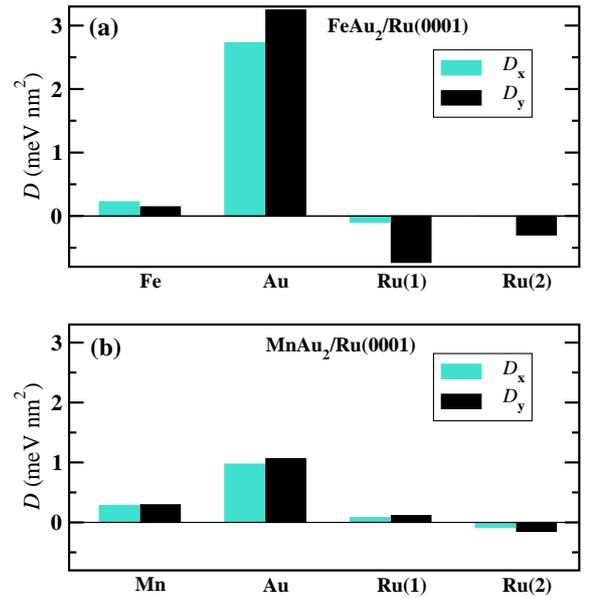}
\caption{(Color online) Atom-wise contributions to non-zero components of $\mathbf{D}$ for (a) FeAu$_2$/Ru(0001) and (b) MnAu$_2$/Ru(0001).
Along [110] $D_y \neq0 = D_x$, while along [1$\overline{1}$0] $D_x \neq0 = D_y$. \textit{X}, Au and Ru($n$) indicate the magnetic atom (Fe or Mn), Au atom 
and $n$-th layer Ru atoms, respectively, where $n$=1 and 2.}
\label{fig:dmi_onRu_layer_wise}
\end{figure}

We have also extracted the contributions to $\bf D$ that arise from each kind of atom,\cite{heidePBCM2009} 
focusing on the \textit{X}, Au, Ru(1) and Ru(2) atoms, as the DM interaction is expected to be significant 
only near the surface. [Note that here, by the contribution of the Ru(2) atoms we mean the average contribution 
of the Ru(2)-I and the Ru(2)-II atoms.] These atom-wise contributions are depicted graphically in 
Fig.~\ref{fig:dmi_onRu_layer_wise}. It is interesting to note that for both FeAu$_2$/Ru(0001) and 
MnAu$_2$/Ru(0001), the largest (positive) contributions arise from the Au atoms. The contributions coming from 
the \textit{X} atoms also always enhance $\mathbf{D}$, but the magnitude is much smaller compared to those from 
the Au atoms. The Ru(1) and Ru(2) atoms have contributions reducing $\mathbf{D}$, except for MnAu$_2$/Ru(0001), 
where Ru(1) contributes additively.  There are two possible ways in which two magnetic atoms can interact through 
the DM mechanism, either directly, or involving a third, ``non-magnetic" atom.\cite{FertJMMM1980} In our case, 
this third atom could be either Ru or Au. Our results suggest that it is this latter, three-site mechanism that 
is dominant in our case. The much larger contribution from Au atoms is in accordance with the large spin-orbit 
coupling in Au and the strong buckling of the overlayer. The larger buckling in FeAu$_2$/Ru(0001) than MnAu$_2$/Ru(0001) 
also leads to a stronger contribution to $\bf D$.

\subsubsection{Results for Magnetic Anisotropy Energy $K$}

The third contribution to the energies of spin spirals on FeAu$_2$/Ru(0001) and MnAu$_2$/Ru(0001)
consists of the magnetic anisotropy energy $K$. For both these systems, we have calculated the 
energy barriers for a rotation of the magnetic moment in the $xz$ and $yz$ planes, which are given respectively by:

\begin{eqnarray}
K{^{110}}  = & E(\theta = \theta^{\rm hard}_1, \varphi=0) - E(\theta=\theta^{\rm easy}_1, \varphi = 0), \nonumber \\\label{eqn:mae_K110}\\
K{^{1\overline{1}0}} = & E(\theta = \theta^{\rm hard}_2, \varphi=\frac{\pi}{2}) - E(\theta=\theta^{\rm easy}_2, \varphi = \frac{\pi}{2}), 
\nonumber \\\label{eqn:mae_K1-10}
\end{eqnarray}

\noindent
where $E$ is the energy, obtained including spin-orbit interactions,  and the moments on the $X$ atoms are
constrained to point along the direction specified by the angles ($\theta$, $\varphi$); the polar angle
$\theta$ is measured from the surface normal and the azimuthal angle $\varphi$ is measured from the $x$-axis 
[see Fig.~\ref{fig:system_image}(a)]. The easy and hard axes for the two rotations are specified by the angles 
$\theta^{\rm easy}_i$ and $\theta^{\rm hard}_i$. $K$ has two contributions: $K_{\rm{SO}}$ and $K_{\rm{dip}}$, 
which arise from spin-orbit interactions, and magnetic dipolar interactions, respectively.

Let us first consider $K_{\rm{SO}}$, which can be calculated in two possible ways: either self-consistently (SC) 
or by using the force theorem (FT). To check the applicability of the FT for the calculation of the MAE of the 
systems considered here, we first compute as a test the quantity 
$K_{\rm test} = E(\theta = \frac{\pi}{2}, \varphi=0) - E(\theta=0, \varphi = 0)$, using both approaches. The number 
of k-points required for a converged  SC calculation is 256 in the full Brillouin zone, while 4096 k-points are needed 
for the FT calculations. For FeAu$_2$/Ru(0001), we obtain $K_{\rm test}$ = 1.14 and 0.93 meV per Fe atom, from the FT and SC 
approaches, respectively; the corresponding values for MnAu$_2$/Ru(0001) are 0.18 meV and 0.19 
meV per Mn atom. Based upon this, we conclude that results for $K_{\rm{SO}}$ using the two approaches are likely 
to agree to the desired degree of accuracy. Henceforth, we have used the FT to calculate all the values of $K$ reported 
in this section. 

We now proceed to vary $\theta$, keeping $\varphi$ fixed at a constant value $\varphi_c$, which is equal to
either 0 or $\frac{\pi}{2}$, when determining $K$ along the [110] and [1$\overline1$0] directions, respectively. 
We define 

\begin{equation} 
E_\perp (\theta, \varphi_c)  = E^{\rm{SO}}(\theta, \varphi_c) -  E^{\rm{SO}}(0, \varphi_c).  \label{eqn:mae_perp}
\end{equation}

\begin{figure} [!h]
\centering
\includegraphics[width=7.0cm ]{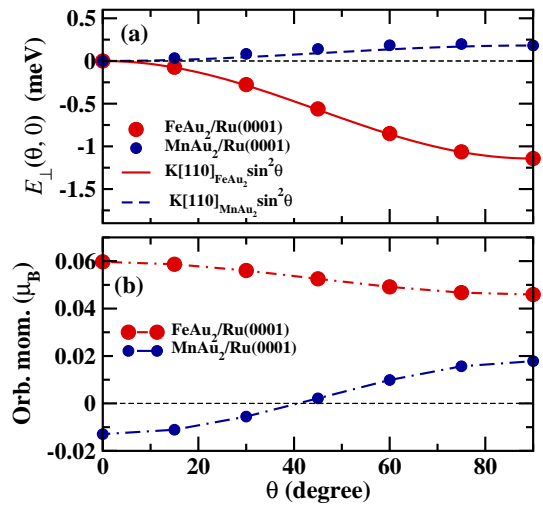}
\caption{(Color online) The variation of (a) $E_\perp (\theta ,  \varphi_c=0)$  and (b) orbital moment of 
the \textit{X} atom as a function of the polar angle $\theta$ for FeAu$_2$/Ru(0001) and MnAu$_2$/Ru(0001). 
The easy axis lies along the direction of minimum $E_\perp (\theta,  \varphi_c=0)$.
} \label{fig:mae_onRu}
\end{figure}

\noindent
In Fig.~\ref{fig:mae_onRu}(a), we have plotted our results for $E_\perp$ for FeAu$_2$/Ru(0001) and 
MnAu$_2$/Ru(0001), with $\varphi_c = 0$. The results for the two systems are quite different. For 
FeAu$_2$/Ru(0001), the highest value of $E_\perp$ occurs for $\theta = 0$, and the lowest value for 
$\theta = \frac{\pi}{2}$, whereas for MnAu$_2$/Ru(0001), the angles corresponding to the highest and 
lowest values of $E_\perp$ are reversed. This suggests that (assuming that the contribution from dipolar 
interactions is small; this remains to be verified below) the position of the hard and easy axes is 
interchanged in the two systems studied here. It is also interesting to note that $K$  is significantly 
higher for FeAu$_2$/Ru(0001) than for MnAu$_2$/Ru(0001).

\begin{table}[!h]
\centering
\caption{Results for magnetic anisotropy energy $K$, along with $K_{\rm{SO}}$ and $K_{\rm{dip}}$, the 
contribution due to spin-orbit coupling  and magnetic dipole-dipole interaction, respectively, along two 
high-symmetry directions, for \textit{X}Au$_2$/Ru(0001).} 
\begin{tabular}{l c r r r}
\hline 
\hline
\textit{X} & direction & $K_{\rm{SO}}$           &  $K_{\rm{dip}}$  & $K$ \\
                          &            & \multicolumn{3}{c}{(meV per \textit{X} atom)}     \\
\hline
Fe  & [110]               &   1.14   &   0.04    &   1.18\\ 
    & [1$\overline{1}$0]  &   1.18   &   0.04    &   1.22\\[0.2cm]
Mn  & [110]               &   0.18   &  $-0.06$  &   0.12\\
    & [1$\overline{1}$0]  &   0.14   &  $-0.06$  &   0.08\\                   
\hline
\hline
\end{tabular}
\label{table:mae}
\end{table}
  
We have also shown, in Fig.~\ref{fig:mae_onRu}(b), how the orbital moment per \textit{X} atom changes as 
$\theta$ is varied. One observes a sinusoidal variation, though the amplitude of variation is small. We 
find that, for both systems, the highest and lowest values of orbital moment occur when the magnetization
is along the hard-axis and the easy-axis, respectively. Note that this contradicts with the 
prediction of Bruno.\cite{brunoPRB1989}  The prediction is based on the assumption that the majority 
and minority $d$-bands are well separated by the exchange interaction; though this is true for the $X$ atoms, 
for the Ru atoms this assumption does not hold true.   

Next, we consider $K_{\rm{dip}}$, which arises from the magnetostatic interaction between the magnetic 
moments. We find that the contributions to $K$ from dipolar interactions are significantly smaller than
those from the spin-orbit interaction, especially for FeAu$_2$/Ru(0001). Our results for $K_{\rm{dip}}$ are
listed in the fourth column of Table ~\ref{table:mae}. The negative sign of $K_{\rm{dip}}$ for the Mn
alloy indicates that the easy axis is out-of-plane here, while dipolar interactions always favor an in-plane
axis for ferromagnetic configurations.

Finally, the total value of $K$ is obtained by adding $K_{\rm{SO}}$ and $K_{\rm{dip}}$ (see the last column
in Table~\ref{table:mae}). We also obtain  $K_{\rm{avg}}$, the average value of $K$ in the (110) and 
(1$\overline{1}$0) planes [see Eq.~(\ref{eqn:energy_total})]; in all the cases studied here, $K_{\rm avg} = {K}/{2}$ .
We find that the easy axis lies in-plane for FeAu$_2$/Ru(0001), but out-of-plane for MnAu$_2$/Ru(0001).

\subsubsection{Results for $E_{\rm{total}}$}

\begin{figure} [!h]
\centering
\subfigure{\label{fig:feau2_onRu_ss_all}\includegraphics[width=8cm]{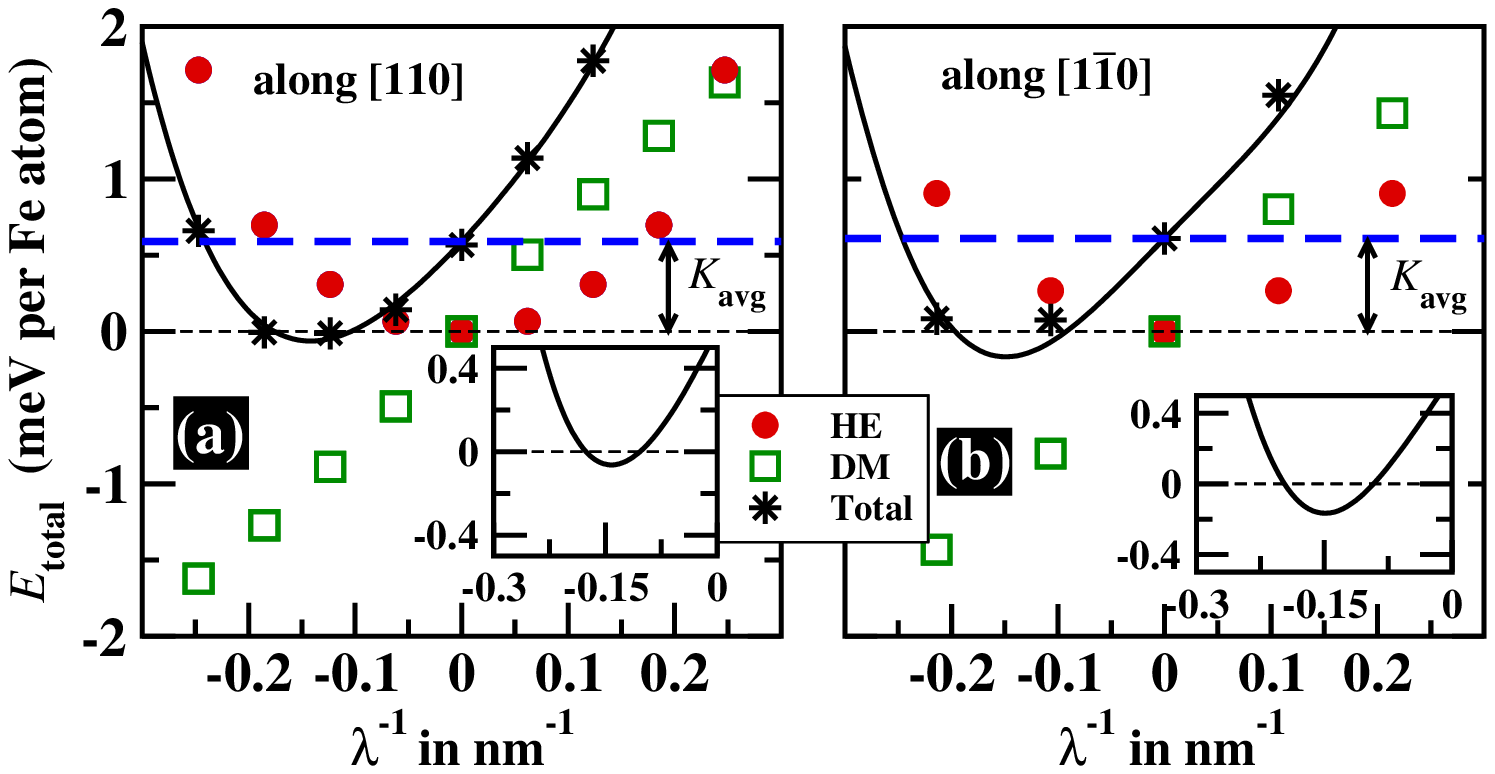}
}
\hspace{2pt}
\subfigure{\label{fig:mnau2_onRu_ss_all}\includegraphics[width=8cm]{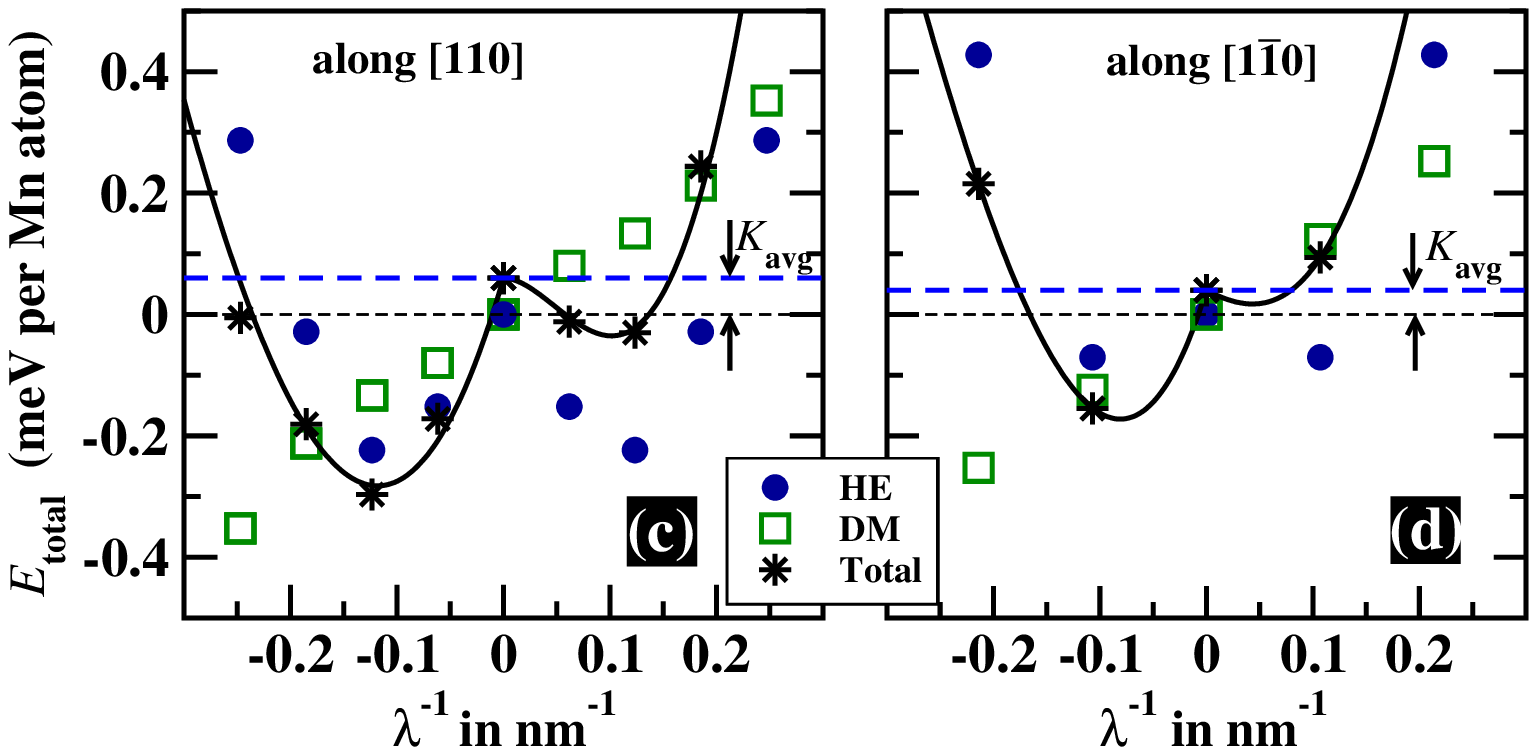}
}\\
\caption{(Color online) Variation of total energy ($ E_{\rm{total}} $) as a function of $\lambda^{-1}$ for  
[(a) and (b)] FeAu$_2$/Ru(0001) and [(c) and (d)]  MnAu$_2$/Ru(0001) along [${110}$] and [$1\overline{1}0$]. The 
insets in (a) and (b) are zoomed in areas using the same scale as in (c) and (d).
} \label{fig:total_ss_onRu}
\end{figure}

Having obtained the values of $E_{\rm{HE}}$, $E_{\rm{DM}}$ and $K_{\rm{avg}}$, we are now finally in a 
position to calculate the total energy $E_{\rm{total}}$ for FeAu$_2$/Ru(0001) and MnAu$_2$/Ru(0001). In 
Fig.~\ref{fig:total_ss_onRu}, the values of $E_{\rm{HE}}$ and $E_{\rm{DM}}$ are shown by circles and 
squares, and the value of $K_{\rm{avg}}$ is shown by dashed lines. The final values $E_{\rm{total}}$, 
obtained by adding these three terms, are shown by the stars and the solid black curves fit to them. 

For FeAu$_2$/Ru(0001), we find that along the [110] direction, the most energetically favorable state 
is a left-rotating spin spiral with $\lambda{^{-1}} = -0.14$ nm$^{-1}$, this is lower in energy than 
the FM state by 0.06 meV per Fe atom [see Fig.~\ref{fig:total_ss_onRu}(a)]. Along [1$\overline{1}$0], a 
spin spiral of $\lambda{^{-1}} = -0.15$ nm$^{-1}$ becomes lower in energy than the FM state by 0.17 meV 
per Fe atom [see Fig.~\ref{fig:total_ss_onRu}(b)]. This latter spin spiral, with a wavelength of 6.7 nm, is 
the lowest-energy magnetic ground state for FeAu$_2$/Ru(0001). However, 
it is only very slightly lower in energy than the FM state.

Similarly, the lower two panels of Fig.~\ref{fig:total_ss_onRu} show the various contributions to the
energies of spin spirals on MnAu$_2$/Ru(0001). We see that a left-rotating spin spiral with 
$\lambda{^{-1}} = -0.12$ nm$^{-1}$ along $[110]$ is lower in energy than the FM state by 0.28 meV per
Mn atom, while a  left-rotating spin spiral with $\lambda{^{-1}} = -0.08$ nm$^{-1}$ along 
$[1\overline{1}0]$ is lower in energy than the FM state by 0.17 meV per Mn atom. Of these, the former, 
with a wavelength of 8.5 nm, is the magnetic ground state.

\subsection{Comparison with Freestanding Alloy Monolayers of \textit{X}Au$_2$}

In order to gauge what effect the Ru substrate has, we now repeat the previous calculations, but
for freestanding \textit{X}Au$_2$ alloy monolayer systems, i.e., in the absence of the Ru substrate.
As before, we present separately the contributions to the total energy from each term in Eq.~(\ref{eqn:energy_total}).

\subsubsection{Results for Heisenberg Exchange Energy $E_{\rm{HE}}$}

We have calculated $E_{\rm{HE}}$ for freestanding FeAu$_2$ and MnAu$_2$ monolayers. From the already
presented  calculations on collinear magnetic structures (see Section IV.A above), we have seen that 
both FeAu$_2$ and MnAu$_2$ freestanding monolayers remain flat upon relaxation, i.e.,  no buckling is 
observed. We now restrict ourselves to calculating 
$E_{\rm{HE}}$ self-consistently (SC) for flat freestanding monolayers (see also the Appendix). We find that it is adequate to use 78 k-points 
in the irreducible Brillouin zone.

\begin{figure} [!h]
\centering
\includegraphics[width=7cm]{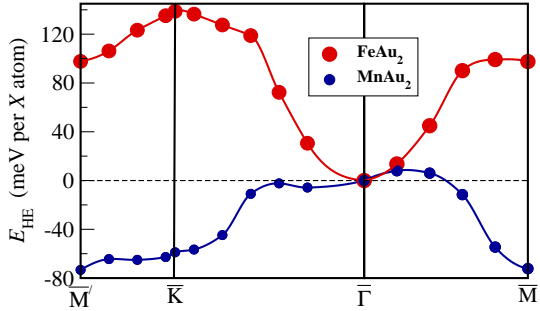}
\caption{(Color online) Energy dispersion $E_{\rm{HE}}$ for freestanding monolayers of FeAu$_2$ [shown by gray (red) 
dots] and MnAu$_2$ [shown by dark (blue) dots]. The ground state magnetic structure for FeAu$_2$ is ferromagnetic 
(minimum at $\overline{\Gamma}$), whereas that for MnAu$_2$  is row-wise antiferromagnetic (minimum at 
$\overline{\mathrm{M}}$).}
\label{fig:ss_uml}
\end{figure}

Fig.~\ref{fig:ss_uml} shows our results for $E_{\rm{HE}}$ along high-symmetry directions of the Brillouin 
zone for both FeAu$_2$ and MnAu$_2$ freestanding monolayers. We find that the lowest $E_{\rm{HE}}$ states 
are FM for FeAu$_2$ -- see the large (red) dots -- and row-wise AFM for MnAu$_2$ -- see the small (blue)
dots. Thus, for the freestanding alloy monolayers, we find that the collinear magnetic states 
are lower in energy than the spin spiral states. Note that for flat monolayers, $E_{\rm{DM}}$ is identically 
zero. So, upon going from freestanding alloy monolayers to surface alloys on Ru(0001), the magnetic ground 
state changes from FM to a spin spiral state in the case of FeAu$_2$/Ru(0001), while for MnAu$_2$/Ru(0001), 
the ground state changes from a row-wise AFM state to a spin spiral state.

\subsubsection{Results for Dzyaloshinskii-Moriya Energy $E_{\rm{DM}}$}

Though the freestanding \textit{X}Au$_2$ monolayers do not show any buckling, in order be able to compare the 
asymmetric exchange coupling between the \textit{X} and Au atoms with and without the substrate, we have obtained 
the value of $\bf{D}$ as a function of the buckling parameter $d_{\rm{b}}$ of the freestanding monolayers of 
\textit{X}Au$_2$. We have taken $d_{\rm{b}}$ to be 0.5, 1.0 and 1.5 \AA. Further, in order to better enable a 
comparison with the corresponding systems on the Ru substrate, we have also considered $d_{\rm{b}}$ to be 0.38 
\AA\ (for the FeAu$_2$ monolayer) and 0.22 \AA\ (for the MnAu$_2$ monolayer); these values correspond to the 
values of $d_{\rm{b}}$ for the overlayer in the \textit{X}Au$_2$/Ru(0001) systems.  

\begin{figure} [!h]
\centering
\includegraphics[width=7.5cm]{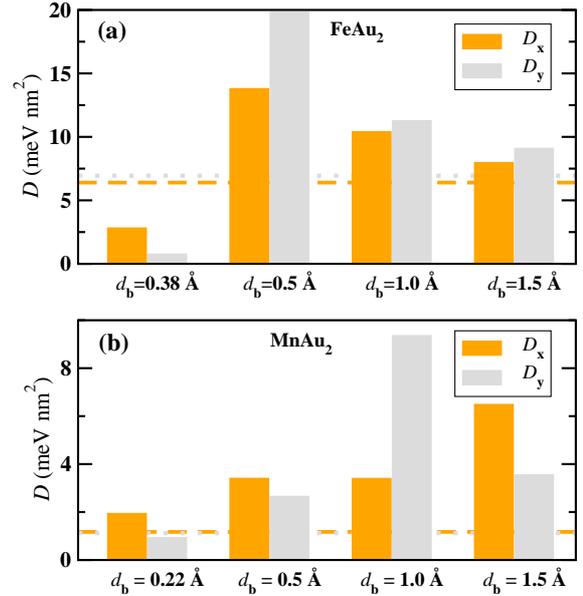}
\caption{(Color online) The variation of $D_x$  and $D_y$ as a function of overlayer buckling 
$d_{\rm{b}}$ for freestanding monolayers of (a) FeAu$_2$  and (b) MnAu$_2$. Along [110] $D_y \neq0 = D_x$, 
while along [1$\overline{1}$0] $D_x \neq0 = D_y$. The dashed and dotted lines correspond 
to the values of $D_x$  and $D_y$, respectively, in the case of FeAu$_2$/Ru(0001) and MnAu$_2$/Ru(0001). 
Note that the $y$-axis scale is different in (a) and (b).}
\label{fig:dmi_uml}
\end{figure}

In Fig.~\ref{fig:dmi_uml}, we have plotted our results for $D_x$  and $D_y$, as a function of buckling parameter 
$d_{\rm{b}}$, for freestanding FeAu$_2$ and MnAu$_2$ monolayers; the component $D_z$ vanishes in all the cases 
considered here. For purposes of comparison, the values of $D_x$ and $D_y$ for the corresponding 
\textit{X}Au$_2$/Ru(0001) systems are shown by dashed and dotted lines, respectively. We find that for FeAu$_2$ 
monolayers, the values of both $D_x$ and $D_y$ at $d_{\rm{b}} = 0.38$ \AA\ are much smaller than the corresponding
values for FeAu$_2$/Ru(0001). In contrast, for MnAu$_2$ monolayers, the values of $D_x$ and $D_y$ are similar to 
the values for MnAu$_2$/Ru(0001). It is therefore difficult to reach any general conclusions about the effect of 
the Ru substrate; it is apparently system-dependent, since the magnetic interactions between the Fe and Ru atoms 
differ from those between the Mn and Ru atoms. We also observe that for the alloy monolayers, the values of 
$D_x$ and $D_y$ can differ a lot, whereas the values are almost the same for the \textit{X}Au$_2$/Ru(0001) systems. 

For FeAu$_2$/Ru(0001) the largest values of $D_x$  and $D_y$ occur for $d_{\rm{b}} = 0.5$ \AA, while for 
MnAu$_2$/Ru(0001), we find the values are the largest at $d_{\rm{b}} = 1.0$ \AA. We obseve that two competing effects 
determine the magnitude of $\mathbf{D}$: (i) a geometrical effect that enhances its value with increasing buckling and 
(ii) the influence of the distance between $X$ and Au, that decreases $\mathbf{D}$ if the $X$-Au distance is too large. 
This is in line with the model of Levy and Fert~\cite{FertJMMM1980} for the DM interaction.

\subsubsection{Results for Magnetic Anisotropy Energy $K$}

We have calculated the value of the magnetic anisotropy energy $K$ for freestanding monolayers of 
\textit{X}Au$_2$. Test calculations show that results obtained using the FT and SC methods are comparable. 
We therefore continue by using the FT. We find that it suffices to use 4096 and 6400 $\mathbf{k}_{\|}$-points in the Brillouin 
zone for the calculation of $K_{\rm{SO}}$ for the FeAu$_2$ and MnAu$_2$ monolayers, respectively. 

\begin{figure} [!h]
\centering
\includegraphics[width=7.0cm ]{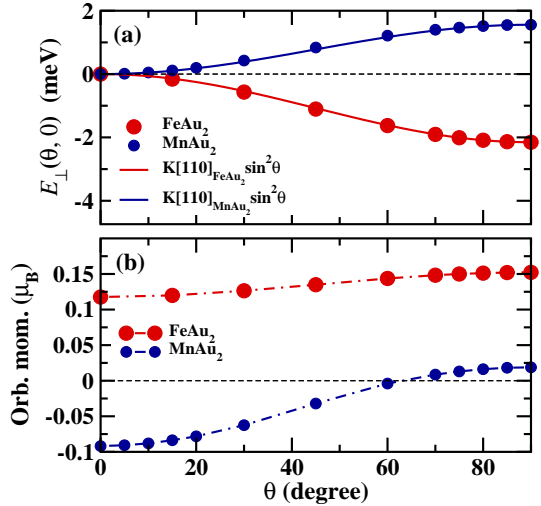}
\caption{(Color online) The variation of (a) $E_\perp (\theta, \varphi_c=0)$  and (b) orbital moment of 
the \textit{X} atom  as a function of the polar angle $\theta$ for FeAu$_2$ and MnAu$_2$ freestanding 
monolayers. The easy axis lies along the direction of minimum $E_\perp (\theta, \varphi_c=0)$.}
\label{fig:mae_uml}
\end{figure}

\begin{table}[!h]
\centering
\caption{Values of $K_{\rm{SO}}$ and $K_{\rm{dip}}$, the contributions to MAE due to spin-orbit coupling  and 
dipole-dipole interaction, respectively, for freestanding \textit{X}Au$_2$ monolayers.
$\mu_z$ and $\mu_x$ are the orbital moments when magnetization points along $z$- and $x$-axis, respectively.} 
\begin{tabular}{l r r r r}
\hline 
\hline
\textit{X} & $K_{\rm{SO}}$ &  $K_{\rm{dip}}$  &  $\mu_z$ & $\mu_x$ \\
                          &  \multicolumn{2}{c}{(meV per \textit{X} atom)}   & \multicolumn{2}{c}{\ \ \ ($\mu_B$)}  \\
\hline
Fe  &    2.1   &   0.05    &  0.12   & 0.15 \\ 
Mn  &    1.8   &  $-0.08$  & $-0.09$ & 0.02 \\                  
\hline
\hline
\end{tabular}
\label{table:mae_uml}
\end{table}

In Fig.~\ref{fig:mae_uml}(a), we have plotted our results for $E_\perp (\theta,  \varphi_c = 0)$ vs.~$\theta$.
The small (red) dots and large (blue) dots show the results for freestanding FeAu$_2$ and 
MnAu$_2$ monolayers, respectively.  We have fitted $E_\perp$ to ${\rm sin}^{2} \theta$ (solid line). The positions 
of the minimum and maximum values of $E_\perp$ are seen to be the same as in the presence of the Ru(0001) substrate, 
and are again opposite for the two different $X$. However, the values of $K_{\rm{SO}}$ are found to have become 
significantly larger in the absence of the Ru substrate.

From Fig.~\ref{fig:mae_uml}(b), we see that the value of the orbital moment per $X$ atom is the highest along 
the easy-axis for the freestanding monolayers of FeAu$_2$ and MnAu$_2$. This behavior differs from the 
trend seen in the \textit{X}Au$_2$/Ru(0001) systems, though it is consistent with the prediction of 
Bruno.\cite{brunoPRB1989}  The values of orbital moments are larger for the freestanding monolayers than 
the corresponding values on the deposited monolayers. This quenching of the orbital moments is an 
effect of the crystal field of the substrate.

We find that the value of $K_{\rm{SO}}$ does not differ appreciably in the [110] and $[1\overline{1}0]$
directions; it is ~2.1 meV per Fe atom for the FeAu$_2$ monolayer, and~1.8 meV per Mn atom for the MnAu$_2$ 
monolayers. Note that the direction of the easy axis differs in the two cases, for the former it is in-plane,
while for the latter it is out-of-plane.

Next, we have obtained the values of $K{_{\rm{dip}}}$  for the freestanding alloy monolayers of 
\textit{X}Au$_2$; once again we do not find an appreciable difference between our results for the [110] and 
$[1\overline{1}0]$ directions. For FeAu$_2$ monolayers the value of $K{_{\rm{dip}}}$ is $0.05$ meV per Fe 
atom, compared to the value of $0.04$ meV per Fe atom for FeAu$_2$/Ru(0001) (see Table~\ref{table:mae}).
The values of $K{_{\rm{dip}}}$ for MnAu$_2$ and MnAu$_2$/Ru(0001) are $-0.08$ and $-0.06$ meV per Mn atom, 
respectively. The slightly higher values of $K{_{\rm{dip}}}$ for the freestanding monolayers arise from 
the higher values of the magnetic moments.

\section{Discussion}

\subsection{Stability of the Surface Alloys}
In Table~\ref{table:delta_H} we had determined the stability of the surface alloys FeAu$_2$/Ru(0001) and 
MnAu$_2$/Ru(0001), with respect to the phase-segregated elemental monolayers on Ru(0001). We had also examined 
the miscibility of the corresponding freestanding monolayers. In this table, however, we had only considered 
collinear magnetic structures for the four surface alloy systems. We have now found that the magnetic ground 
state for the two supported surface alloy systems is not the ferromagnetic state but a spin spiral; however,
the energy difference between these two is small. Therefore, the values in Table~\ref{table:delta_H} do not 
change appreciably upon considering non-collinear configurations for the mixed phases. For FeAu$_2$/Ru(0001), 
$\Delta H$ is further lowered by only 0.04\% further, while for MnAu$_2$/Ru(0001), $\Delta H$ is further 
decreased by only 0.05\%.

\subsection{Relative Contributions of Different Terms to $E_{\rm{total}}$}

It is interesting to note that the primary reason for obtaining the spin spiral ground states is different 
for FeAu$_2$/Ru(0001) and MnAu$_2$/Ru(0001). By examining Fig.~\ref{fig:total_ss_onRu}, we see that for 
FeAu$_2$/Ru(0001), it is the DM interaction that is chiefly responsible for stabilizing the spin spiral 
ground state over the FM state. In contrast, for  MnAu$_2$/Ru(0001) the predominant role is played by 
the Heisenberg exchange interaction. One reason why the DM interaction is stronger in the case of 
FeAu$_2$/Ru(0001) is the larger value of the buckling parameter $d_{\rm{b}}$ in this system. From comparison 
to the unsupported alloys it can be seen that, in addition, also the chemical nature of the magnetic element 
and its modification by the Ru substrate by bonding (charge transfer) have an important influence on the strength 
of this interaction. The propensity of Heisenberg exchange to favor spin spirals is indicated by a small value of 
$\Delta E_{\rm{AFM-FM}}$; we have already seen above that this quantity is smaller for MnAu$_2$/Ru(0001) than 
FeAu$_2$/Ru(0001).

\subsection{Role of Au}

It was mentioned earlier that Fe/Ru(0001) shows a 120$^{\circ}$ N\'{e}el structure\cite{hardratPRB2009} 
and Mn/Ru(0001) shows a row-wise AFM structure.\cite{dupuisPRB1993} In this study, we see that the magnetic 
interaction changes in these systems, due to the presence of Au in the overlayer, and the systems are driven 
toward a spin spiral ground state in FeAu$_2$/Ru(0001) and MnAu$_2$/Ru(0001), due to complex magnetic 
interactions.  From Section IV.C2, we also see that due to the large spin-orbit coupling constant of Au atoms, 
the DM interaction mainly acts via these atoms, rather than the Ru atoms.
On the other hand, it is the Ru substrate that makes the Au contribution very different in the Fe and
Mn systems. The large additive contribution of 
the Au atoms toward the DM parameter thus helps to lower the energy of a spin spiral compared to the FM state, 
especially in the case of FeAu$_2$/Ru(0001).

\section{Summary}

In summary, we have performed \textit{ab initio} density functional theory calculations to obtain the magnetic 
ground states of two surface alloys: FeAu$_2$/Ru(0001) and MnAu$_2$/Ru(0001). By considering both collinear and 
non-collinear magnetic structures, we have found that the magnetic ground state for both systems corresponds to 
a left-rotating spin spiral.  For the Fe system the spiral propagates along the $[110]$ direction with a period 
of 6.7~nm, while in the Mn alloy it is along $[1\overline{1}0]$ and has a period of 8.5~nm. In the former case, 
the spin spiral is stabilized by the Dzyaloshinskii-Moriya interaction, whereas in the latter case it is primarily 
stabilized by the Heisenberg exchange interaction. These results show that magnetic surface alloys constitute a 
new class of systems that can be explored for the existence of spin spirals. However, in the two particular 
systems considered in this work, the spin spiral states are only slightly lower in energy than the ferromagnetic 
state, by 0.17 and 0.28 meV per Fe and Mn atom, respectively. 

We have seen that the strength of the DM interactions is very sensitive to the buckling of the overlayer. Of the 
two surface alloy systems considered in this study, FeAu$_2$/Ru(0001) has a buckling that is almost twice as large 
as that observed in MnAu$_2$/Ru(0001),  and the values of $D_x$ and $D_y$ are larger by a factor of $\sim$6--7. For 
a monolayer on a substrate, the buckling is fixed, being determined by the size mismatch between the overlayer and 
substrate atoms. However, when one considers surface alloys of the type AB/C, as in this study, one has more 
parameters to play with, since the buckling depends not only on the size difference between the overlayer atoms and 
the substrate, but also on the size difference between the two overlayer constituents A and B. One therefore has the 
ability to tune the buckling, and thus the strength of the DM interaction, over a wider range; this can be made use 
of as a way of further stabilizing spin spirals. 

We also find that FeAu$_2$/Ru(0001) has a significantly high magnetic anisotropy energy, of the order of 1 meV per Fe atom, 
with an in-plane easy-axis. On the other hand, MnAu$_2$/Ru(0001) has a magnetic anisotropy energy that is smaller by an order 
of magnitude, and an out-of-plane easy axis. Upon comparing these values of the MAE with those obtained for the corresponding 
freestanding monolayers, we find that for both the systems, the presence of the substrate does not alter the direction of 
the easy-axis, but reduces the magnitude of the MAE considerably.    

It has been shown in earlier work that magnetic interactions are primarily responsible for stabilizing the FeAu$_2$/Ru(0001)
surface alloy against phase-segregation.\cite{mehendalePRL2010} In addition to confirming these results, we now find that this 
is also true for the MnAu$_2$/Ru(0001) system, as well as the corresponding freestanding monolayers. For  FeAu alloys on Ru(0001), 
it has been concluded that an important role is played by ferromagnetically polarized substrate Ru atoms.\cite{marathePRB2013} 
However,  these are of course absent in the freestanding alloy monolayers, while even in our two supported alloy systems, we 
find that the substrate Ru atoms are primarily spin polarized antiparallel to the overlayer Mn atoms. The main driving force 
for alloy formation is that by forming an alloy structure where the ``magnetic" (Fe or Mn) atom is surrounded in the overlayer 
by Au atoms, it can raise its magnetic moment, and thus significantly lower the exchange energy. This effect is sufficiently 
strong to flip the enthalpy of formation from being positive to being negative.

By comparing with the corresponding freestanding alloy monolayers, we find that the presence of the Ru substrate plays 
a crucial role in determining the magnetic properties of the surface alloy systems and Au atoms in the overlayer promote 
chirality in these systems.

Our results underline the need for considerable caution when applying the magnetic force theorem in calculations of magnetic 
structures, when small energy scales are involved. 

Most importantly, we wish to underline that our work shows that bimetallic magnetic surface alloy systems of the type AB/C, 
such as those studied here, allow one to play with and tune the Dzyaloshinskii-Moriya interaction, thus allowing one to access 
novel magnetic structures such as spin spirals. Such surface alloys, which contain a heavy atom in the topmost layer, give one 
a way to  the DM interaction via the structure, in contrast to A/B thin-film systems. This leads to the possibility of manipulating 
the spin spiral via electric fields or adsorbates.

\appendix
\section{Applicability of the Force Theorem}

In section IV.C1, we showed that the force theorem failed to yield an accurate value for $\delta E_{\rm{HE}}$,
the difference in energy from Heisenberg exchange interactions, between two spin spirals of slightly different 
wavelength, especially for MnAu$_2$/Ru(0001). Here, we discuss this issue further. 

We first consider conical spin spirals, where the magnetic moments are not constrained to lie in a plane, but 
precess around the axis of rotation, making an angle $\alpha$ with it. Note that for planar spin spirals, 
$\alpha= \frac{\pi}{2}$. The energy difference  $\delta E{_{\rm{HE}}}(\alpha)$ between two spin spirals with 
wavevectors $\mathbf q_1$ and $\mathbf q_2$ can be written as:\cite{lezaicUnpublished}

\begin{eqnarray} \label{eqn:delta_E_cone}
\delta E{_{\rm{HE}}}(\alpha) & = & E{_{\rm{HE}}^{\mathbf q_1}}(\alpha) - E{_{\rm{HE}}^{\mathbf q_2}}(\alpha),\label{eqn:delta_E_cone_1} \\
 & \simeq & {\rm{sin}}^{2}\alpha \ ( E{_{\rm{HE}}^{\mathbf q_1}}(\dfrac{\pi}{2}) - E{_{\rm{HE}}^{\mathbf q_2}}(\dfrac{\pi}{2})), \nonumber\\
& \equiv & {\rm{sin}}^{2}\alpha \ \delta E{_{\rm{HE}}}(\dfrac{\pi}{2}) , \label{eqn:delta_E_cone_2}
\end{eqnarray} 

\noindent
where the approximation holds for a small difference between  $\mathbf q_1$ and $\mathbf q_2$, and does not depend on 
the method of calculation (FT or SC). For our calculations we have taken $|\mathbf q_1|$ and $|\mathbf q_2|$  to be 
0 and 3.3 nm$^{-1}$, respectively, for MnAu$_2$/Ru(0001), and 8.8 and 11 nm$^{-1}$, respectively, for freestanding MnAu$_2$, 
along the [110] direction. 

\begin{figure} [!h]
\centering
\includegraphics[width=7.0cm ]{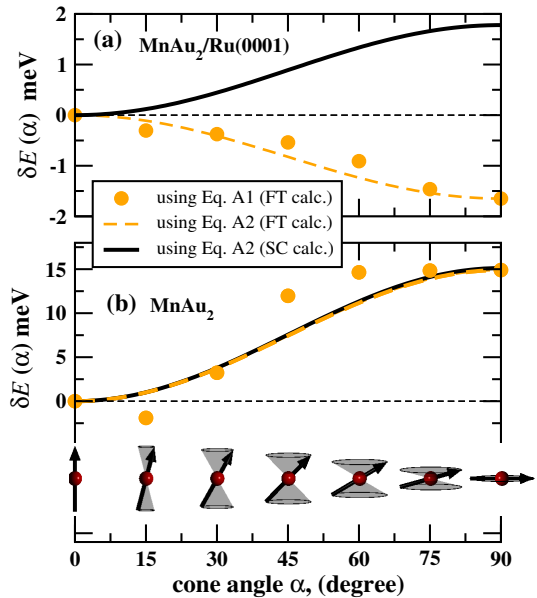}
\caption{(Color online) The variation of $\delta E{_{\rm{HE}}}(\alpha)$ as a function of the cone angle $\alpha$ 
for (a) MnAu$_2$/Ru(0001) and (b) MnAu$_2$ freestanding monolayers. The gray (orange online) dots and dashed 
lines are obtained from FT calculations, while the black line is obtained from SC calculations. The bottommost row of 
figures show how the spin precesses around the axis of rotation as the cone angle $\alpha$ is varied. 
}
\label{fig:mnau2_Ru_cone_angle}
\end{figure}

First, we checked whether the approximation holds in our case. In order to do this, we have obtained $\delta E{_{\rm{HE}}}(\alpha)$ 
from Eqs.~(\ref{eqn:delta_E_cone_1}) and (\ref{eqn:delta_E_cone_2}) by using the FT for different values of $\alpha$.  
We have used 5041 $\mathbf{k}_{\|}$-points in the irreducible Brillouin zone of MnAu$_2$/Ru(0001), while for the freestanding 
monolayer of MnAu$_2$, 2304 $\mathbf{k}_{\|}$-points are used. The results thus obtained are shown in Fig.~\ref{fig:mnau2_Ru_cone_angle}, 
where we have plotted $\delta E{_{\rm{HE}}(\alpha)}$ as a function of $\alpha$ for both MnAu$_2$/Ru(0001) and a 
freestanding MnAu$_2$ monolayer. The values obtained from Eqs.~(\ref{eqn:delta_E_cone_1}) and (\ref{eqn:delta_E_cone_2}) 
are shown by  dots and  dashed lines, respectively. For both the systems, we see that the values obtained from 
Eq.~(\ref{eqn:delta_E_cone_1}) deviate slightly from sinusoidal behavior, but agree qualitatively with the values obtained 
from Eq.~(\ref{eqn:delta_E_cone_2}). 

Having shown that values of $\delta E{_{\rm{HE}}}$ can be calculated approximately by the FT method from 
Eq.~(\ref{eqn:delta_E_cone_2}), we proceed to calculate the values self-consistently (SC) using only 
Eq.~(\ref{eqn:delta_E_cone_2}).  The number of k-points used for these calculations is 800 and 400 for 
MnAu$_2$/Ru(0001) and freestanding MnAu$_2$, respectively. By comparing the energies obtained for MnAu$_2$/Ru(0001) 
from SC (black solid line in Fig.~\ref{fig:mnau2_Ru_cone_angle}) and FT (dots and dashed line) methods, we see 
that the values of $ \delta E{_{\rm{HE}}}(\alpha)$ do not agree with each other, in fact, they even have opposite 
sign. On the other hand, for the freestanding monolayer of MnAu$_2$, the values match quite well. This suggests 
that the FT is applicable for the calculation of $\delta E{_{\rm{HE}}}$  between two spin spirals (both conical 
and planar) of the freestanding MnAu$_2$ monolayer, at least when the difference in wavevectors is small. However, 
the FT breaks down for MnAu$_2$/Ru(0001) even when $\alpha$ is very small. We conclude that the applicability of the 
force theorem has to be tested for all cases individually, since the breakdown of the FT for MnAu$_2$/Ru(0001) could 
not have been anticipated, either from the comparison to the freestanding layers, or to the FeAu$_2$/Ru(0001) case.

\section*{Acknowledgement}
\noindent
Useful discussions with S.~Rousset, B.~Zimmermann, M.~Marathe, B.~Schweflinghaus and Ph.~Mavropoulos are gratefully acknowledged. 
S.~Biswas acknowledges CSIR, India for a research fellowship and J\"{u}lich Supercomputing Centre (JSC) for computation time.

\end{document}